\documentclass[epj,nopacs]{svjour}
\usepackage{fullpage}
\usepackage{graphicx,graphics,tabularx,wrapfig,subfigure,citesort,footmisc,amsmath,verbatim,calc,epsfig}
\usepackage[sort&compress]{natbib}
\usepackage[normalem]{ulem}
\usepackage{setspace}
\usepackage{amssymb}
\usepackage[usenames,dvipsnames]{color}
\usepackage{xr}
\usepackage{xspace}
\bibpunct{[}{]}{,}{n}{}{}

\newcommand{\MBNExplorer}{\textsc{MBN Explorer}\xspace}

\begin{document}

\title{
Studying chemical reactions in biological systems with MBN Explorer:
implementation of molecular mechanics with dynamical topology
}

\author{
Gennady B. Sushko\inst{1,2} \and
Ilia A. Solov'yov\inst{3,4} \and
Alexey V. Verkhovtsev\inst{1,2} \and
Sergey N. Volkov\inst{5} \and
Andrey V. Solov'yov\inst{2,4}
}

\institute{
Goethe-Universit\"{a}t Frankfurt am Main, Max-von-Laue-Str. 1, 60438 Frankfurt am Main, Germany
\and
MBN Research Center, Altenh\"{o}ferallee 3, 60438 Frankfurt am Main, Germany
\and
University of Southern Denmark (SDU), Campusvej 55, 5230, Odense M, Denmark
\and
On leave from A.F. Ioffe Physical-Technical Institute, Polytekhnicheskaya ul. 26, 194021 St. Petersburg, Russia
\and
Bogolyubov Institute for Theoretical Physics, Metrolohichna str. 14-b, Kiev, 03680, Ukraine
}

\date{}

\abstract{
The concept of molecular mechanics force field has been widely accepted nowadays for studying
various processes in biomolecular systems. In this paper, we suggest a modification for the standard
CHARMM force field that permits simulations of systems with dynamically changing molecular topologies.
The implementation of the modified force field was carried out in the popular program \MBNExplorer,
and, to support the development, we provide several illustrative case studies where dynamical topology
is necessary. In particular, it is shown that the modified molecular mechanics force field can be applied
for studying processes where rupture of chemical bonds plays an essential role, e.g., in irradiation- or
collision-induced damage, and also in transformation and fragmentation processes involving biomolecular systems.
}

\maketitle

%%%%%%%%%%%%%%%%%%%%%%%%%%%%%%%%%%%%%%%%%%%%%%%%%%%%%%%
\section{Introduction}
\label{Introduction}

Nowadays, it has become feasible to study structure and dynamics of molecular systems that constitute
of millions of atoms~\cite{sanbonmatsu2007high,zhao2013mature} and evolve on time scales up to hundreds
of nanoseconds~\cite{salomon2013routine} by employing the classical molecular mechanics (MM) approach.
In this approach, a molecular system is treated classically, so that constituent atoms interact with
each other through a parametric phenomenological potential that is governed by the type of individual
atoms and by the network of chemical bonds between them. This network defines a so-called molecular topology,
that is a set of rules that impose constraints in the system and permit maintaining its natural shape,
mechanical, and thermodynamical properties.
The MM method has been widely used throughout the last decades
\cite{rappe1997molecular, gumbart2012mechanisms, shim2013detection, zhao2013mature} and implemented in
the well-established computational packages, such as CHARMM~\cite{CHARMMProgram}, AMBER~\cite{AMBERprogram},
GROMACS~\cite{GROMACSprogram}, and NAMD~\cite{NAMD}.

\MBNExplorer~(www.mbnexplorer.com)~\cite{MBN_ExplorerPaper} is an alternative, emerging software for the
simulation of complex biomolecular, nano- and mesoscopic systems. It is suitable for classical molecular dynamics (MD),
Monte Carlo~\cite{dick2010nanoparticles, dick2011fragmentation, JCC:JCC23613,solov2014thermally,solov2013simulation}
and relativistic dynamics simulations~\cite{Sushko2013404,sushko2013sub,sushko2013simulations,sushko2015multi}
of a large range of molecular systems, such as nano-~\cite{refId0,Verkhovtsev201380} and biological systems,
nanostructured materials~\cite{sushko2014validation,ISolovyov08},
composite/hybrid materials~\cite{solov2009possibility,geng2009uncovering,geng2010fullerene,moskovkin2014simulation},
gases, liquids, solids and various interfaces~\cite{sushko2014molecular, verkhovtsev2013molecular},
with the sizes ranging from atomic to mesoscopic.
Among other applications, \MBNExplorer can be used to simulate thermo-mechanical damage of a biological medium,
e.g. a DNA nucleosome, which is caused by the propagation of a shock wave initiated by irradiation with
fast ions~\cite{Yakubovich2012}. The results of such simulations are used then to evaluate the efficiency of
radiation with different projectiles~\cite{Surdutovich2013} within the framework of the multiscale approach
to the physics of radiation damage~\cite{Surdutovich2014} and can be applied in the field of ion-beam cancer
therapy~\cite{Solovyov2009_IBCT, Surdutovich2014, Surdutovich2010_EPJD, Surdutovich2013}.

Despite numerous successes, the conventional MM method is primarily capable of studying processes where
chemical reactions do not take place. This leads to significant limitations of the method and makes it
practically unsuitable for studying highly non-equilibrium processes in biomolecular systems,
e.g. thermo-mechanical biodamage. This particular example involves rupture and formation of covalent bonds
that cannot be simulated by the conventional MM method due to a fixed topology of the system.

Simulation of the rupture and formation of covalent bonds can be performed by using
Quantum Mechanical/Molecular Mechanical (QM/MM) me\-thods or {\it ab initio} MD simulations
\cite{QM-MM_review, CPMD_1985_PhysRevLett.55.2471, AIMD_Marx_Hutter_2000}.
Both methods are computationally rather demanding and, thus,
the {\it ab initio} approach is used typically for studying fragmentation of small
biomo\-lecules, such as DNA nucleobases or nucleotides~\cite{Kohanoff_2012_JACS.134.9122, Kohanoff_2015_JPCL.6.3091}.
The size of such systems is far from the typical sizes of systems of biological relevance,
consisting of hundred thousands of atoms, and more. This problem is addressed to some
extent in QM/MM methods where a core part of a large biomolecular system is described
quantum mechanically while all the surroundings are described classically using,
for example, the conventional MM method~\cite{Luedemann_2015_JACS.137.1147, Sjulstok_2015_SciRep}.
Thus, the rupture or formation of covalent bonds can be simulated only in a small part
of the system, which is treated quantum mechanically.

In this paper,
we present an extended version of the conventional MM method, which has been recently implemented in
\MBNExplorer, and de\-monstrate that this extension describes correctly the dynamically changing
molecular topology of a system within the classical MD framework.
The presented modification takes into account additional parameters of the system, such as dissociation
energy of bonds, bonds multiplicity and the valence of atoms. The functional form of the interatomic
interactions is also adjusted to account for the finite dissociation energy of the chemical bonds.
Finally, three examples of simulations with the extended MM method are presented
to demonstrate a proof of principle for utilizing the force field with dynamic topology.
The first two case studies illustrate the processes of rupture and formation of covalent bonds
in a small biomolecule, namely an alanine dipeptide,
which is one of the simplest building blocks of larger biomolecular systems like polypeptides or proteins.
Having proven the force field to work on a simple dipeptide allows us to generalize the framework towards
macromolecules.
This will allow for studying the systems of biologically relevant sizes, on the time scales which are not
accessible by means of {\it ab initio} methods.
The last example illustrates the process of water splitting and the evolution of chemical equilibrium.
It is demonstrated that the results of the simulation are in a reasonable quantitative agreement with
those of the analytical calculations.

%%%%%%%%%%%%%%%%%%%%%%%%
\section{Theoretical approach}
\label{Physical model}

\subsection{Molecular mechanics potential}

The MM potential represents a phenomenological parametrization of the potential energy of a system
and is widely used to describe structure and properties of macromolecular systems, such as
polypeptides~\cite{beck2005cutoff, solov2008alpha,yakubovich2008theory, yakubovich2009phase, piana2014assessing},
proteins \cite{henriques2008rational, solov2012decrypting, solov2014separation, barragan2014identification, piana2014assessing},
DNA \cite{hart2011optimization, zou2011recognition, volkov2012micromechanics, Surdutovich2013},
lipids~\cite{klauda2010update, pastor2011development,barragan2014identification,skjevik2015all},
and many others.
All physically important interactions in a system, i.e. both covalent and non-bonded long-range interactions,
are accounted for using a simple parametric form, so that the total energy of the system reads as:
\begin{equation}
U_{\rm tot} = U_{\rm cov} + U_{\rm vdW} + U_{\rm Coul} \ .
\label{Utot_MM}
\end{equation}

\noindent
The terms on the right-hand side describe the covalent, van der Waals, and
electrostatic Coulomb interactions, respectively.
The van der Waals interaction between two neutral atoms or molecules is usually
modeled by the Lennard-Jones (LJ) potential:
\begin{equation}
%U_{\rm vdW} \approx
U_{\rm LJ} = \epsilon
\left[
\left( \frac{r_0}{r_{ij}} \right)^{12}  - 2 \left( \frac{r_0}{r_{ij}} \right)^{6}
\right] \ ,
\end{equation}

\noindent
where $\epsilon$ is the depth of the potential energy well,
$r_0$ is the equilibrium distance, and $r_{ij}$ is the distance between atoms.
The function $U_{\rm cov}$ parameterizes the covalent interactions through
a number of empirical parameters and fitting functions.
\MBNExplorer implies the following parametrization for the $U_{\rm cov}$ term:
\begin{eqnarray}
\nonumber
U_{\rm cov} =
\sum_{\substack{\alpha=1  \\  i,j \in \alpha}}^{N_{\rm b}}
U_{ij}^{{\rm (bond)}} +
\sum_{\substack{\alpha=1  \\  i,j,k \in \alpha}}^{N_{\rm a}}
U_{ijk}^{{\rm (angle)}} + \\
+
\sum_{\substack{\alpha=1  \\  i,j,k,l \in \alpha}}^{N_{\rm d}}
U_{ijkl}^{{\rm (dihedral)}} +
\sum_{\substack{\alpha=1  \\  i,j,k,l \in \alpha}}^{N_{\rm i}}
U_{ijkl}^{{\rm (improper)}} \ .
\label{MM_potential}
\end{eqnarray}

\begin{figure}[t]
\centering
\resizebox{0.9\columnwidth}{!}{\includegraphics{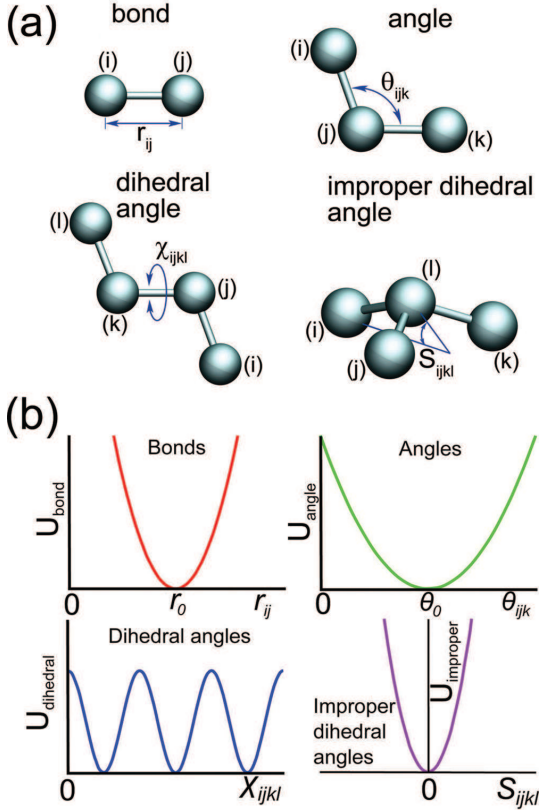}}
%\resizebox{0.9\columnwidth}{!}{\includegraphics{Figures/Fig1_lo-res.eps}}
\caption{(a) Internal coordinates describing molecular mechanics interactions:
$r_{ij}$ governs bond stretching, $\theta_{ijk}$ represents the angular term,
$\chi_{ijkl}$ gives the dihedral angle, and the small out-of-plane angle $S_{ijkl}$
is governed by the so-called improper dihedral angle.
(b) Dependencies of the potential energy on coordinates used in the molecular
mechanics potential, Eq.~(\ref{MM_potential}), describing the bonded, angular,
dihedral angular, and improper dihedral angular interactions.}
\label{MM_fig01}
\end{figure}

\noindent
It is important to stress here that MM potential is only applicable to the systems with a
predefined topology, i.e. a set of rules that define chemical bonds in the system.
The energy of the covalent interactions is then modeled as a sum over all such interactions.
The first and the second terms on the right-hand side of Eq.~(\ref{MM_potential}) describe
the components of the potential energy of the system arising due to stretching of the bonds
between two atoms and due to variation of angles between every topologically defined triplet
of atoms, respectively (see Figure~\ref{MM_fig01}a).
The third term is known as the torsion energy, which is characterized through a dihedral angle
formed by every four atoms connected via covalent chemical bonds. The last term describes the
so-called improper dihedral angles that are used in the molecular topology to maintain planarity.

A widely used MM potential, called CHARMM \cite{CHARMM}, represents a set of parameters
for the simulation of structure and dynamics of bio-/macro\-molecular systems.
This force field employs harmonic approximation for describing the interatomic interactions,
thereby limiting its applicability to small deformations of the molecular system.
The form of the potential energy functions, which are used to describe the components of
$U_{\rm cov}$ in the standard CHARMM force field, is demonstrated in Figure~\ref{MM_fig01}b.
In the case of substantial deformations, the interaction forces should decrease to zero as
the valence bonds rupture. The rupture of valence bonds should also cause the involved
angular and dihedral interactions to vanish as well.
In this work, we report on advances in the \MBNExplorer development, that permit classical
MD simulations of the rupture of covalent bonds by using a dissociative CHARMM potential.
This approach goes beyond the harmonic approximation, thus describing the physics of molecular
dissociation more accurately, and permits construction of dynamic molecular topology, which
instructs \MBNExplorer how the existing covalent bonds can break and new covalent bonds can be formed.
These features make \MBNExplorer rather unique, for example, for simulating irradiation- and
collision-induced biodamage by means of classical MD.
To the best of our knowledge, presently there is no similar approach implemented in other
software for classical MD simulations and, therefore, the implemented algorithms are seen as
unique know-how of \MBNExplorer.
The adopted methodology is described in detail further in this section.

\subsection{Rupture of covalent bonds}

In order to model rupture of covalent bonds in the CHARMM
force field, \MBNExplorer uses modified interaction potentials
which describe the interactions of atoms connected by chemical bonds.
The standard CHARMM force field treats the covalent interactions within the
harmonic approximation as
\begin{equation}
U^{{\rm (bond)}}_{ij} = k_{ij}^{{\rm b}} (r_{ij} - r_0)^{2} \ .
\label{eq:harmBonds}
\end{equation}

\noindent
Here $k_{ij}^{{\rm b}}$ is the force constant of the bond stretching,
$r_{ij}$ is the distance between atoms $i$ and $j$, and the parameter
$r_0$ is the equilibrium covalent bond length.
The above parametrization describes well the bond stretching regime in the case of
small deviations from $r_0$ but gives an erroneous result for larger distortions.
For a satisfactory description of the covalent bond rupture it is reasonable to
substitute the parabolic potential with the Morse potential.
It requires one additional parameter, as compared to the harmonic approximation~(\ref{eq:harmBonds}),
and takes into account the energy of bond dissociation.
Potential energy of the system of two atoms interacting via the Morse potential reads as:
\begin{equation}
U_{\rm M}(r_{ij}) = D_{ij}
\left[ e^{-2\beta_{ij}(r_{ij}-r_0)}-2 e^{-\beta_{ij}(r_{ij}-r_0)} \right] \ ,
\label{eq:MorsePot}
\end{equation}

\noindent
where $D_{ij}$ is the bond dissociation energy and
the parameter $\beta_{ij}$ determines steepness of the potential.
It follows from Eq.~(\ref{eq:MorsePot}) that $U_{\rm M}(r_0)=-D_{ij}$.

Let us consider a small deformation of the covalent bond from its equilibrium
distance, $r_{ij} - r_0 \ll r_0$.
In this case, the energy of the bond can be approximated harmonically as:
\begin{equation}
U_{M}(r_{ij}) \approx - D_{ij} + {\beta_{ij}}^{2}D_{ij}(r_{ij}-r_0)^{2}k \ ,
\label{eq:MorseApprox}
\end{equation}

\noindent
so that
\begin{equation}
\beta_{ij} = \sqrt{k_{ij}^{{\rm b}}/D_{ij}} \ .
\label{eq:betaDef}
\end{equation}

\noindent
This expression defines $\beta_{ij}$ for a certain covalent bond and relates
it to the value of $k_{ij}^{{\rm b}}$ used in the standard CHARMM force field.

\begin{figure}[t]
\centering
\resizebox{0.95\columnwidth}{!}{\includegraphics{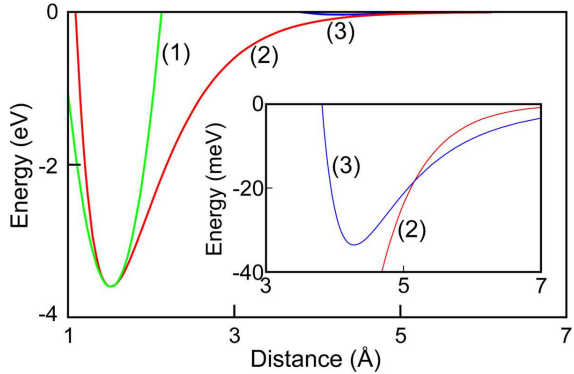}}
%\resizebox{0.95\columnwidth}{!}{\includegraphics{Figures/Fig2_lo-res.eps}}
\caption{
The pairwise carbon-carbon (CHARMM type CN7~--~CN8B) interaction potential in harmonic (1)
and Morse approximations (2). The van der Waals interaction between the two atoms is
illustrated with the blue line denoted as (3).}
\label{fig:Morse_vs_Harmony}
\end{figure}

Figure~\ref{fig:Morse_vs_Harmony} illustrates the Morse potential which models the interaction
of two carbon atoms, having the CN7~--~CN8B type according to the CHARMM nomenclature.
This covalent bond occurs, for example, between the C4'-C5' atoms of ribose.
At small deviations from $r_0$, the Morse potential (red curve) and the harmonic approximation
(green curve) are close to each other.
With increasing interatomic distance ($r_{ij} \gg r_0$), the atoms start to
interact through polarization forces modeled through the Lennard-Jones potential.
The comparison between the Morse and the Lennard-Jones potentials at larger
distances is shown in the inset.
It follows that both potentials are close to each other at the distances of about the
van der Waals contact distance for the non-bonded interactions, which is about 2~\AA~for
the considered carbon atoms.

\subsection{Rupture of valence angles}

The rupture of chemical bonds in the course of simulation automatically employs an improved
potential for the valence angles. In the CHARMM force field, the potential associated with
the change of a valence angle between bonds with indices $ij$ and $jk$ reads as:
\begin{equation}
U^{\rm (angle)}_{ijk} = k^{\rm a}_{ijk} \left(\theta_{ijk} - \theta_{0}\right)^{2} \ ,
\label{eq:angleHarmonic}
\end{equation}

\noindent
where $k^{\rm a}_{ijk}$ and $ \theta_{0}$ are parameters of the potential,
and $\theta_{ijk}$ is the actual value of the angle formed by the three atoms.
This potential grows rapidly with increasing the angle, and it may lead
to non-physical results when modeling the covalent bond rupture.
In order to avoid such cases, in the modified force field the harmonic potential~(\ref{eq:angleHarmonic})
is substituted with an alternative parametrization, which reads as:
\begin{equation}
U^{\rm (cos)}_{ijk} = 2k^{\rm a}_{ijk} \left[ 1 - \cos(\theta_{ijk}-\theta_{0}) \right] \ .
\label{eq:AngleCos}
\end{equation}

\noindent
At small variations of the valence angle, this para\-metrization is identical to the harmonic
approximation (\ref{eq:angleHarmonic}) used in the standard CHARMM force field.
For larger values of the angle, the new parametrization~(\ref{eq:AngleCos}) defines an energy
threshold which becomes important for an accurate modeling of bond breakage.

\begin{figure}[t]
\centering
%\resizebox{0.95\columnwidth}{!}{\includegraphics{Figures/Fig3.eps}}
\resizebox{0.95\columnwidth}{!}{\includegraphics{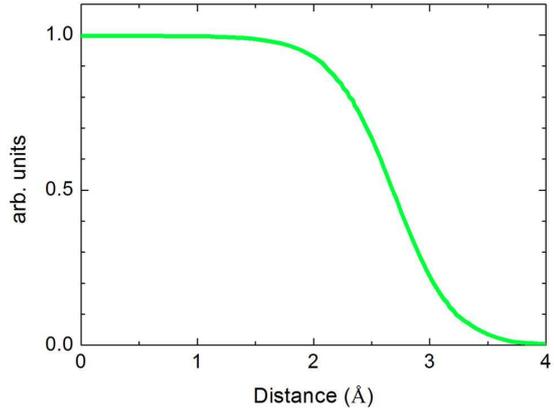}}
\caption{The switching function $\sigma(r_{ij})$ calculated for the carbon-carbon (CHARMM type CA~--CA)
interaction using Eq.~(\ref{eq:sigma}). The function is used to rupture angular interactions in the
modified CHARMM force field.}
\label{fig:sigmaFunc}
\end{figure}

The rupture of a covalent bond is accompanied by a rupture of the angular interactions
associated with this bond. The effect of bond breakage on the angular potential can be
described through a special function $\sigma(r_{ij})$, defined as
\begin{equation}
\sigma(r_{ij})= \frac{1}{2}
\left\{ 1 - \tanh \left[ \beta_{ij}(r_{ij} - r_{ij}^*) \right] \right\}  \ ,
\label{eq:sigma}
\end{equation}

\noindent
with $r_{ij}^*=(R_{ij}^{\rm vdW}+ r_{0})/2$.
Figure~\ref{fig:sigmaFunc} illustrates that $\sigma(r_{ij})$ has a form of a smoothed step function.
This function introduces a correction to the angular interaction potential, assuming that
the distance between two atoms involved in an angular interaction increases from the equilibrium value,
$r_{0}$, up to the van der Waals contact value, $R_{ij}^{\rm vdW}$. Since an angular interaction
depends on two bonds connecting the atoms with indices $ij$ and $ik$, the potential energy, describing
the valence angular interaction that is subject to rupture, is parameterized as
\begin{equation}
\tilde{U}^{\rm (angle)}_{ijk} = \sigma(r_{ij}) \, \sigma(r_{jk}) \, U^{\rm (cos)}_{ijk} \ .
\label{eq:angleRupture}
\end{equation}

\begin{figure}[t]
\centering
\resizebox{0.92\columnwidth}{!}{\includegraphics{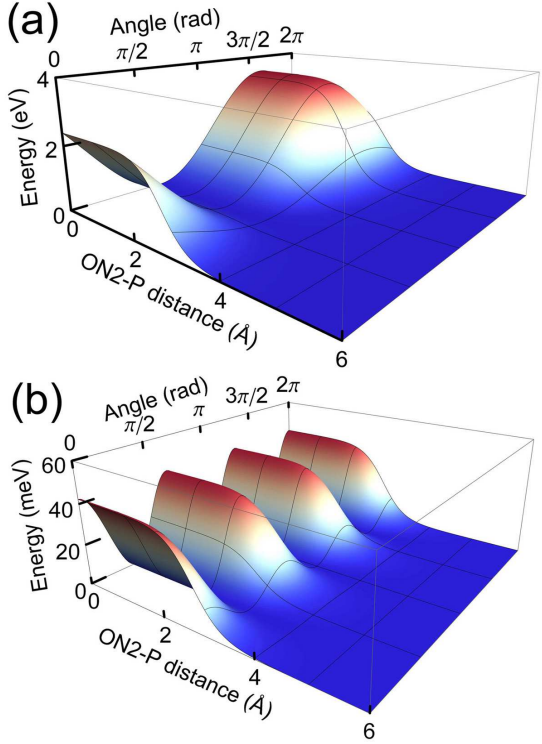}}
%\resizebox{0.92\columnwidth}{!}{\includegraphics{Figures/Fig4_lo-res.eps}}
\caption{(a) The CN8B~--~ON2~--~P angular potential calculated using Eq.~(\ref{eq:angleRupture}) with
account for the ON2~--~P bond rupture. (b) The CN4~--~P~--~ON2~--~CN7 dihedral potential calculated
using Eq.~(\ref{eq:dihedralRupture}) with account for the ON2~--~P bond rupture.}
\label{fig:OPAngularBreak}
\end{figure}

\noindent
As seen from this expression, the angular potential decreases with the increase of the bond length between
any of the two pairs of atoms $ij$ or $jk$. For exemplary purposes, in Figure~\ref{fig:OPAngularBreak}a we
show the CN8B~--~ON2~--~P angular potential, which arises, for instance, when modeling DNA nucleotides.
The presented angular potential was calculated using Eq.~(\ref{eq:angleRupture}) assuming the breakage of the
bond between the oxygen and the phosphorous atoms. For the sake of illustration, the CN8B~--~ON2 bond length
in this case was taken equal to its equilibrium value~$r_{0}$.

\subsection{Rupture of dihedral interactions}

``Dihedral'' interactions arise in the conventional MM potential due to the change
of the dihedral angles between every four topologically defined atoms.
Let us consider a quadruple of atoms with indices $i$, $j$, $k$ and $l$ (see Figure~\ref{MM_fig01}),
bound through an interaction which is governed by a change of the dihedral angle.
In this case, the dihedral angle stands for the angle between the plane, formed by the atoms $i$, $j$ and $k$,
and the plane, formed by the atoms $j$, $k$ and $l$.
In the harmonic approximation, the dihedral-energy contribution reads as:
\begin{equation}
U^{\rm (dihedral)}_{ijkl} = k_{ijkl}^{\rm d}
\left[ 1 + \cos(n_{ijkl} \, \chi_{ijkl} - \delta_{ijkl}) \right] \ ,
\label{eq:Udihedral}
\end{equation}

\noindent
where $k_{ijkl}^{\rm d}$, $n_{ijkl}$ and $\delta_{ijkl}$ are parameters of the potential,
and $\chi_{ijkl}$ is the angle between the planes formed by atoms $i$, $j$, $k$ and $j$, $k$, $l$.

The dihedral interactions also become dis\-turbed upon covalent bond rupture; therefore,
Eq.~(\ref{eq:Udihedral}) should be modified to properly account for this effect.
The rupture of a dihedral interaction between a quadruple of atoms $i$, $j$, $k$ and $l$
should take into account three bonds that contribute to this interaction.
Thus, the potential energy describing the dihedral interaction with account for the bond rupture
reads as:
\begin{equation}
\tilde{U}^{\rm (dihedral)}_{ijkl} =
\sigma(r_{ij}) \, \sigma(r_{jk}) \, \sigma(r_{kl}) \, U^{\rm (dihedral)}_{ijkl} \ ,
\label{eq:dihedralRupture}
\end{equation}

\noindent
where $U^{\rm (dihedral)}_{ijkl}$ is the potential~(\ref{eq:Udihedral}) describing the dihedral
interaction within the framework of the standard CHARMM force field.
The functions $\sigma(r_{ij})$, $\sigma(r_{jk})$, and $\sigma(r_{kl})$ are defined by Eq.~(\ref{eq:sigma});
they are used to limit the dihedral interaction upon increasing the corresponding bond length.
Figure~\ref{fig:OPAngularBreak}b shows a typical profile of a dihedral potential with accounting for the
bond rupture. In this case, we have considered the CN4~--~P~--~ON2~--~CN7 dihedral interaction where the
middle ON2~--~P bond was broken. This interaction is also important when modeling bond breakages in DNA nucleotides.

\subsection{Formation of new bonds}

Rupture of covalent bonds leads to formation of individual atoms, radicals or smaller molecular fragments.
In order to properly simulate the chemical balance in the system, one should allow for the formation of
new bonds.

In \MBNExplorer, after the rupture of a chemical bond between two atoms, these atoms are placed in
a special list of chemically active atoms.
Only the atoms from this list can participate in chemical reactions and form new bonds.
For each atom from this list, the number of possible molecular bonds is stored and determined by its valence.

In order to create new bonds in the system, the list of chemically active atoms is evaluated
at each simulation step, and the neighboring atoms are selected.
A chemical bond is formed between a pair of atoms provided that the following conditions have been met:
(i) the parameters of the bond,
e.g., the equilibrium distance and the value of bond formation energy
for this combination of atoms are defined in the simulation input,
(ii) atoms are modeled as bound through the Morse potential, and
(iii) the distance between the atoms is less than the predefined cutoff/capture radius,
which is an independent parameter for each bond type used to speed-up the simulations.
If all these conditions are met simultaneously in the simulation, the bond is created and the
system's topology is updated.
For each new covalent bond, the neighboring atoms are analyzed.
If the parameters of angular bond are defined for some group of atoms such bond should also be formed.

\subsection{Redistribution of partial charges}

\begin{figure*}[!htb]
\centering
\resizebox{1.8\columnwidth}{!}{\includegraphics{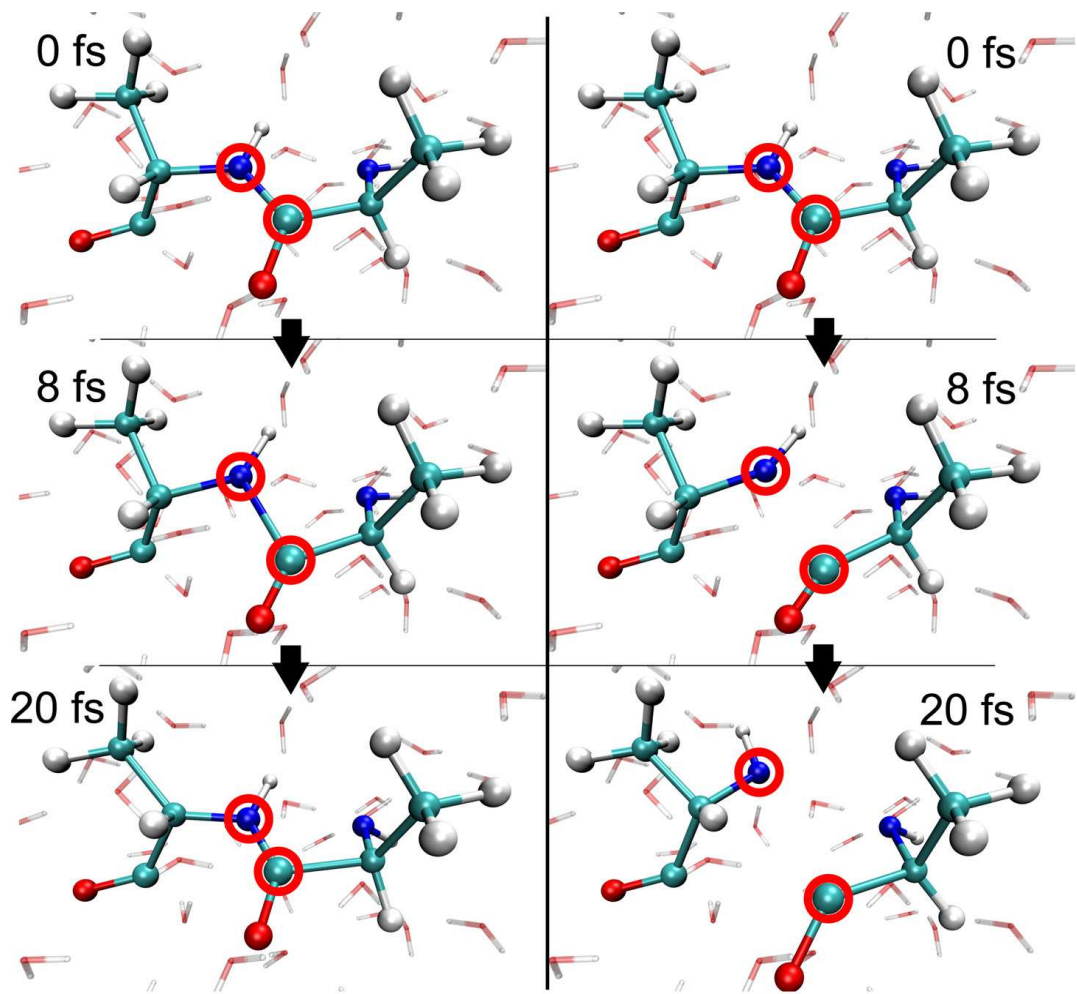}}
%\resizebox{1.8\columnwidth}{!}{\includegraphics{Figures/Fig5_lo-res.eps}}
\caption{Snapshots illustrating dynamics of alanine dipeptide and the C--N bond rupture simulated
with the harmonic~(left) and Morse~(right) potentials at 0~fs~(top), 8~fs~(middle) and 20~fs~(bottom).}
\label{Shapshots}
\end{figure*}

The rupture and formation of bonds leads to reconfiguration of molecular systems and to redistribution
of partial charges of atoms.
The modification of \MBNExplorer presented in this paper accounts for this phenomenon.
The charge redistribution should obey the following conditions: the total charge of the systems is conserved,
the total charge of each individual molecule is an integer in atomic system of units, i.e. an integer number
of the electron charge.
\MBNExplorer supports the two options for modeling the charge redistribution process:
(i) a general (default) one applicable to any molecule, and
(ii) a special one where charges within molecules are redistributed according to the known
electronic configurations.

The default mechanism of charge redistribution is activated upon rupture or formation of covalent bonds.
In the case of a bond rupture, two newly emerged fragments of a molecular system have likely non-compensated,
non-integer charges.
The total charge of each of the newly emerged fragment is thus rounded to the closest integer value and
the charge difference is transferred from one fragment to another.
This difference is redistributed evenly among all atoms of the fragments. Upon the formation of new a bond,
the charge is redistributed inside the newly created molecule in order to lower the values of partial charges
preserving the initial sum of charges.

%%%%%%%%%%%%%%%%%%%%%%%%%%%%%%%%%%%%%%%%%%%%%%%%%%%%%%%%%%%%%%%%%
\section{Numerical Results}
\label{results}

Let us now consider the case studies that go beyond the standard MM methodology.
The first example illustrates the rupture of a single C~--~N bond in an alanine dipeptide molecule.
In the second example, we simulate the reverse process of the new bond formation.
These two case studies demonstrate a proof of principle for the modified MM force field
to describe the dynamically changing molecular topology of the system within the classical MD framework.
%Alanine dipeptide was chosen as a case study because it is one of the simplest building
%blocks of large biomolecular systems like polypeptides or proteins.
%Thus, having proven the force field to work on a small biomolecule, this framework can be
%generalized towards macromolecu\-les.
%
The third example is devoted to the investigation of the process of water splitting at high temperatures.
In this case, the breakage of chemical bonds and the formation of new ones lead to establishing
the chemical equilibrium in the system.

\subsection{Fragmentation of alanine dipeptide}

To illustrate the bond breakage, we have simulated the dynamics of alanine dipeptide
consisting of 20 atoms, solvated in a simulation box with 95 water molecules.
The alanine dipeptide molecule was considered with neutral terminals.

In order to clearly illustrate the difference between the standard CHARMM force field,
utilizing the harmonic interatomic potential, and the dissociative CHARMM potential
implemented in \MBNExplorer, two simulations were carried out.
In these simulations, the rupture of the central C--N bond in the dipeptide, leading
to the formation of two isolated alanine molecules, was monitored.
To facilitate the process, we have set the initial velocity of the C and N atoms high,
which corresponds to an energy fluctuation sufficient for the bond rupture.
In the simulation performed with the standard force field, the peptide bond is modeled
through the harmonic potential, therefore, the bond cannot break.
The behavior of the C--N bond in the harmonic approximation is illustrated in the left
part of Figure~\ref{Shapshots}, and the corresponding atoms are marked with red circles.
In this case, the distance between the atoms oscillates around the equilibrium value
as the atoms always return to their equilibrium positions.

\begin{figure}[!htb]
\centering
%\resizebox{0.95\columnwidth}{!}{\includegraphics{Figures/Fig6.eps}}
\resizebox{0.95\columnwidth}{!}{\includegraphics{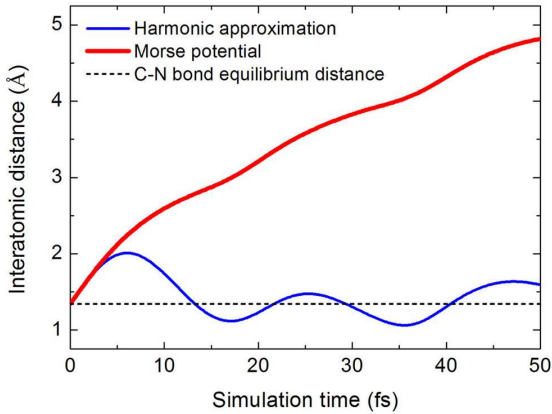}}
\caption{Dependence of the C--N interactomic distance in alanine dipeptide as a function
of the simulation time.}
\label{Chap9_BB}
\end{figure}

In the second simulation, the Morse potential~(\ref{eq:MorsePot}) was used for the
description of the peptide bond.
In this case the C and N atoms do not oscillate around an equilibrium position, and
the structure of the system after 20~fs of simulation changes significantly from the
one considered above (see the right part of Figure~\ref{Shapshots}).
It is evident from the snapshots that the distance between the atoms increases already after 8~fs.

When the distance between the atoms exceeds a given cutoff radius
(which is equal to 2.5~\AA~in this example), the bond is considered as broken.
Once this has happened, the carbon and the nitrogen atoms remain interacting only via
the electrostatic potential and the van der Waals interactions, so that the two alanine
molecules can diffuse apart.
The charge redistribution does not happen in this case because both new fragments of
the dipeptide were initially neutral.

Figure~\ref{Chap9_BB} shows the interatomic distance between the carbon and the nitrogen atoms
as a function of the simulation time.
The equilibrium distance between the atoms is $r_0^{\rm C-N}$ = 1.354~\AA~(dashed line).
The figure demonstrates that in the case of the simulation with the Morse potential,
the interatomic distance monotonically increases indicating that the bond is broken and that
two isolated alanine molecu\-les drift apart.

\subsection{Binding of two alanine amino acids}

\begin{figure}[h]
\centering
%\resizebox{0.95\columnwidth}{!}{\includegraphics{Figures/Fig7.eps}}
\resizebox{0.95\columnwidth}{!}{\includegraphics{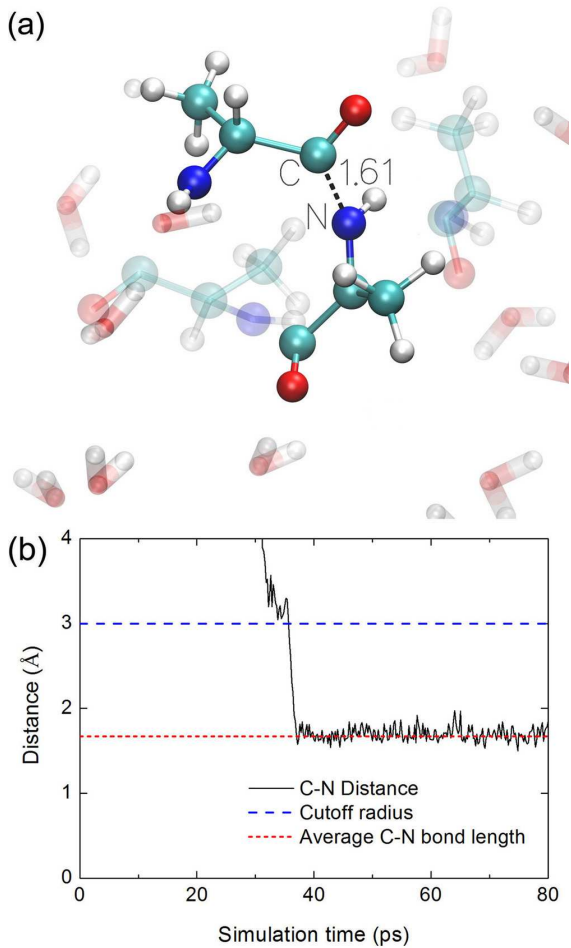}}
\caption{(a) Two alanine molecules approaching each other to form a new C--N bond.
(b) Dependence of the distance between C and N atoms for the two alanine molecules.}
\label{ala-bond}
\end{figure}

The second example illustrates the process of binding two alanine molecules together into
a single dipeptide through the formation of a new covalent bond in the molecular system.
In this case study, six isolated alanine amino acids surrounded by 54 water molecules were
placed in a small simulation box of $24 \times 24 \times 24$~\AA$^3$ with periodic boundary conditions,
and the dynamics of the system was simulated for 80~ps at a fixed temperature of 1000~K controlled
by the Langevin thermostat with the damping time constant of 1~fs.
Each alanine molecule was modeled with unsaturated N-- and C-- termini, i.e. having two unpaired
chemical bonds.

In the course of the simulation, all distances between the different termini of alanines were monitored.
When the distance between a pair of terminal atoms became smaller than the predefined cutoff radius
(equal to 3~\AA~in this example), a new covalent bond was considered to be formed.
Figure~\ref{ala-bond}a illustrates a spacial conformation of two amino acids in the simulation leading
to the formation of a new bond.
Figure~\ref{ala-bond}b gives the dependence of the distance between C and N atoms for the two molecules
shown in the upper part.
At some point, this distance becomes smaller than the cutoff radius~(blue dashed line), and the two
molecules become connected.
Note that after 40~ps the distance between the C and N atoms oscillates around a constant value
corresponding to the C--N bond equilibrium length. %of $r_0 = 1.66$~\AA.
Since six alanines are considered in this simulation, more of them could self-assemble in a polypeptide
chain but this would require longer simulation.
In this system, each initial alanine mole\-cule has a total charge equal to zero.
Therefore, after the formation of a new molecule the charge redistribution step was not necessary.

\subsection{Water splitting}

The third case study concerns the process of water splitting.
At elevated temperatures, water mol\-ecules can dissociate forming different molecular products.
In such a system, the following reactions take place:
\begin{eqnarray}
%{\rm H}_2{\rm O} & {\rm \rightleftharpoons} & {\rm OH} + {\rm H} \label{eq.reqctions1}\\
%{\rm OH} &{\rm \rightleftharpoons} & {\rm O} + {\rm H} \\
{\rm H}_2{\rm O} & {\rm \rightleftharpoons} & {\rm OH}^{-q_1} + {\rm H}^{+q_1} \label{eq.reqctions1} \\
{\rm OH} &{\rm \rightleftharpoons} & {\rm O}^{-q_2} + {\rm H}^{+q_2} \label{eq.reqctions2} \\
{\rm H}_2 &{\rm \rightleftharpoons}& {\rm H} + {\rm H} \label{eq.reqctions3} \\
{\rm O}_2 &{\rm \rightleftharpoons}& {\rm O} + {\rm O} \ , \label{eq.reqctions4}
\end{eqnarray}

\noindent
which describe both the mechanisms of rupture and formation of new chemical bonds.
The superscripts $+q_i$ and $-q_i$ ($i = 1,2$) indicate that the reaction products
in Eqs. (\ref{eq.reqctions1}) and (\ref{eq.reqctions2}) carry some non-zero charge.
The principle of detailed balance states that for a reversible process at the equilibrium
a direct process should be equilibrated by its reverse process.
This forms the equilibrium concentrations of reacting species which depend on the temperature
of the medium.
In order to determine the equilibrium concentrations of molecules in our system, one should use
the equations for the equilibrium concentrations of different molecular species \cite{ihara1978feasibility},
which are based on the Dalton's law and the law of mass action~\cite{LL5,atkins2006physical}:
\begin{eqnarray}
\frac{X_{\rm H} X_{\rm OH} P}{X_{\rm H_2O}} &=& \exp\left(-\frac{\Delta G_1}{R T}\right) \label{eq.balance.1}\\
\frac{X_{\rm H} X_{\rm O} P}{X_{\rm OH}} &=& \exp\left(-\frac{\Delta G_2}{R T}\right)\\
\frac{X_{\rm H}^2 P}{X_{{\rm H}_2}} &=& \exp\left(-\frac{\Delta G_3}{R T}\right) \\
\frac{X_{\rm O}^2 P}{X_{\rm O_2}} &=& \exp\left(-\frac{\Delta G_4}{R T}\right) \ ,
\end{eqnarray}

\noindent
where $X_{\rm H}$, $X_{\rm OH}$, $X_{\rm O}$ and $X_{\rm H_2O}$ are the relative concentrations
of atoms and molecules of different types noted by the subscripts,
$\Delta G_i$ are Gibbs energy values -- the dissociation energies of the chemical reactions in
Eqs.~(\ref{eq.reqctions1})-(\ref{eq.reqctions4}),
$P$ is the total pressure in the system, $R$ is the universal gas constant, and $T$ is the temperature.
Taking into account that the total number of hydrogen atoms in the studied system is twice the number
of oxygen atoms, one obtains:
\begin{eqnarray}
\nonumber
2 X_{\rm H_2} + X_{\rm H} + X_{\rm OH} + 2 X_{\rm H_2O} \\
= 2(2X_{\rm O_2} + X_{\rm OH} + X_{\rm O} + X_{\rm H_2O}) \ .
\end{eqnarray}

\noindent
The normalization condition for the partial concentrations/densities can be written accordingly as:
\begin{eqnarray}
\nonumber
X_{\rm H_2} + X_{\rm O_2} + X_{\rm H} \\
+ X_{\rm OH} + X_{\rm O} + X_{\rm H_2O} &=& 1 \ .
\label{eq.balance.2}
\end{eqnarray}

\noindent
In this case study, we have simulated water splitting in a box of
$130\times130\times130$~{\AA$^3$} with periodic boundary conditions.
The studied system consists of 10917 water mol\-ecules randomly placed inside this box.
Simulations for different temperatures were carried out using the Langevin thermostat
with the damping time of 10~fs. The total simulation time was set to 500~ps.
We note that accounting for the dissociation and formation of new bonds in the MM method
with dynamic topology does not affect significantly the performance of the simulation.
The computational cost of the calculation is about the same as for the conventional MD
simulation with the standard CHARMM force field; the simulation of water splitting
was carried out using a 12-core AMD workstation in 3 days.
Due to the constant volume and density, and different temperatures at each simulation,
the pressure in the box was changing from 2500 to 4000~bar which corresponds to the water
vapor state with a high density.
The simulated values of pressure and temperature were used in the analytic expressions,
Eqs.~(\ref{eq.balance.1})-(\ref{eq.balance.2}), for estimating the equilibrium concentrations
of the reaction products.
The values of $G_1 = G_2 = 117.9$~kcal/mol, $G_3 = 104$~kcal/mol, $G_4 = 119$ kcal/mol were used.

\begin{figure*}[t]
\centering
\resizebox{1.8\columnwidth}{!}{\includegraphics{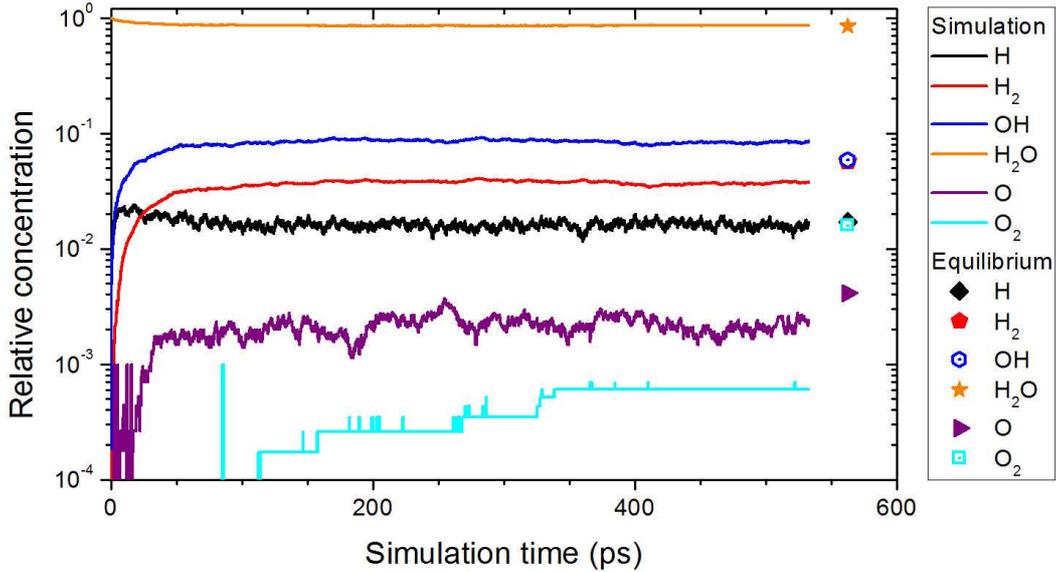}}
%\resizebox{1.8\columnwidth}{!}{\includegraphics{Figures/Fig8_lo-res.eps}}
\caption{Time evolution of the relative concentration of atoms $X_{\rm H}$, $X_{\rm O}$
and molecules $X_{\rm H_2}$, $X_{\rm OH}$, $X_{\rm H_2O}$ in the simulated system at $T=4800$~K.
Results of the simulation are shown with solid lines, dots show corresponding analytic values
of concentrations. }
\label{water-split-t}
\end{figure*}

\begin{figure}[t]
\centering
%\resizebox{0.95\columnwidth}{!}{\includegraphics{Figures/Fig9.eps}}
\resizebox{0.95\columnwidth}{!}{\includegraphics{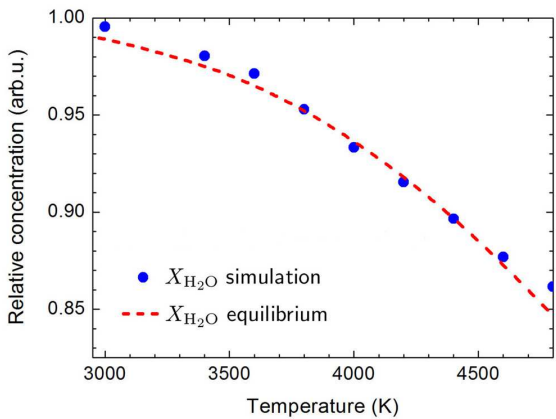}}
\caption{Comparison of the simulated dependence of the relative concentration of water molecules,
$X_{\rm H_2O}$, on temperature with the one obtained from the equilibrium analysis,
Eqs.~(\ref{eq.balance.1})-(\ref{eq.balance.2}).}
\label{water-split-xh2o}
\end{figure}

For each reaction (\ref{eq.reqctions1})--(\ref{eq.reqctions2}), the distribution of
partial charges $q_i$ was calculated separately and set explicitly for all products.
For water, the partial charge of oxygen and hydrogen atoms was set to $-0.834$ and $0.417$,
respectively, for a $\rm OH$ molecule the values of $-0.375$ and $0.375$ were assumed.
${\rm H}_2$ and ${\rm O}_2$ molecules in reactions (\ref{eq.reqctions3}) and (\ref{eq.reqctions4})
were considered as neutral with zero partial charges.

Figure~\ref{water-split-t} shows the time evolution of the relative concentration of atoms and molecules
in the system at $T=4800$~K.
Solid lines present the results of the simulations, while symbols show the results of numerical solution
of Eqs.~(\ref{eq.balance.1})-(\ref{eq.balance.2}).

Initially, all molecules in the system are water molecules. Frequent random collisions lead to their
dissociation into H and OH. During the first 10~ps, these are the two main reaction products and their
relative concentration coincide. After 10~ps the concentration of hydrogen atoms becomes significant and
the reaction ${\rm H} + {\rm H} \to {\rm H}_2$ becomes probable. At this instance the concentration
of ${\rm H}_2$ increases significantly and the concentration of ${\rm H}$ drops. After 100~ps of simulation,
the concentrations of ${\rm H}$, ${\rm H}_2$, ${\rm OH}$ and ${\rm H_2O}$ become nearly constant and close
to the equilibrium values obtained from Eqs.~(\ref{eq.balance.1})-(\ref{eq.balance.2}). However due to the
low concentration of oxygen atoms the formation of ${\rm O}_2$ molecules is a relatively slow process and
requires significant time which we could not reach in our simulation. This process starts after about 100~ps
and does not reach the saturation stage by the end of the simulation.

To further characterize the interconversion of chemical reactions in the system,
in Figure~\ref{water-split-xh2o} we present the comparison of the simulated temperature dependence
of the relative concentration of water mol\-ecules, $X_{\rm H_2O}$, with
the one obtained from the equilibrium analysis, Eqs.~(\ref{eq.balance.1})-(\ref{eq.balance.2}).
The increase of temperature leads to the splitting of water molecules and their number decreases.
The comparison shows that the two curves show practically identical behavior.
Small deviations can be attribu\-ted to the limited simulation time and to the high pressure in the system,
as both of these factors lead to the decrease of the number of split molecules.

\begin{figure}[t]
\centering
%\resizebox{0.95\columnwidth}{!}{\includegraphics{Figures/Fig10.eps}}
\resizebox{0.95\columnwidth}{!}{\includegraphics{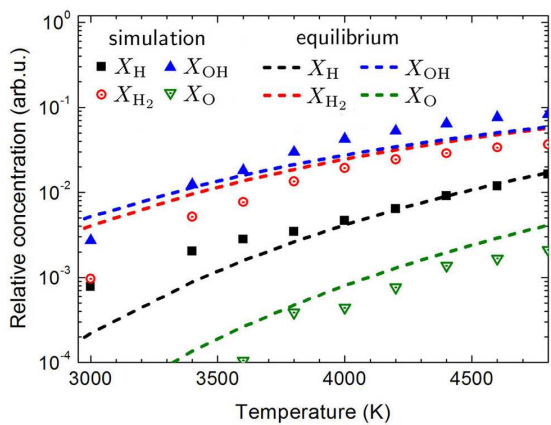}}
\caption{
Comparison of simulated dependencies of $X_{\rm H}$, $X_{\rm H_2}$, $X_{\rm OH}$,
$X_{\rm O}$ on temperature with those obtained from the equilibrium analysis,
Eqs.~(\ref{eq.balance.1})-(\ref{eq.balance.2}).
Dashed lines correspond to the results of the equilibrium analysis,
solid curves correspond to the results of simulations.}
\label{water-split-x}
\end{figure}

A similar analysis was performed for $X_{\rm H}$, $X_{\rm H_2}$, $X_{\rm OH}$, and $X_{\rm O}$.
Figure~\ref{water-split-x} illustrates the dependence of relative concentration on temperature
for the different products of dissociation.
These dependencies are compared with those obtained from the equilibrium analysis based on
Eqs.~(\ref{eq.balance.1})-(\ref{eq.balance.2}).
The figure demonstrates a reasonable agreement of the two approaches, especially at high
temperatures where the simulated values $X_{\rm H}$ nearly coincide with those obtained from
the equilibrium analysis, and the values of $X_{{\rm H}_2}$ and $X_{\rm OH}$ are rather close.
For smaller temperatures, the deviation of the curves increases, primarily because of the insufficient
simulation time.
At higher temperatures, the molecules  diffuse faster, thus allowing to achieve sooner the chemical
equilibrium in the system.

\section{Conclusions}

This paper reports on an important development of the \MBNExplorer software package,
namely on the implementation of reactive force fields based on the modified CHARMM force field.
This modification allows one to simulate and analyze chemical reactions and transformations
involving covalent bond breakages in bio- and organic molecules and molecular systems
\cite{deVries_2002_JPB.35.4373,Zhou_2012_JPCA.116.9217}.
This is obviously a very large research domain and we believe that the implementations reported
in this work will open many new possibilities for the research analysis bas\-ed on the computational
modeling of the mentioned systems.
In this paper, we have not tried to explore all the possible applications of the new computational tool,
but instead focused on the most characteristic and illustrative examples.

This work can be extended further in many different ways. The introduced reactive force fields can be
applied systematically to different systems and the results of simulations can be compared with the
appropriate experiments \cite{deVries_2002_JPB.35.4373,Zhou_2012_JPCA.116.9217}.
The simulation results can also be validated through the comparison with the corresponding results
obtained through the quantum MD simulations of relatively small molecular systems
\cite{Kohanoff_2012_JACS.134.9122, Kohanoff_2015_JPCL.6.3091}.
These and many more other possible studies based on the reported key implementation are the subject
for further research.

%%%%%%%%%%%%%%%%%%
\section*{Acknowledgments}

The possibility to perform computer simulations at the Frankfurt Center for Scientific Computing
is gratefully acknowledged.
The authors acknowledge supercomputer time on Stampede provided the Texas Advanced Computing Center (TACC)
at the University of Texas at Austin through Extreme Science and Engineering Discovery Environment (XSEDE)
Grant XSEDE MCB-120160 (to IAS).
IAS is grateful for the financial support from the Lundbeck Foundation and to the Russian Scientific Foundation
(Grant No. 14-12-00342).

\bibliographystyle{nature}
\bibliography{bib}

\end{document}